\documentclass[%
 reprint,
nofootinbib,
 amsmath,amssymb,
 aps,
]{revtex4-2}

\usepackage{graphicx}
\usepackage{dcolumn}
\usepackage{bm}
\usepackage{slashed}
\usepackage{hyperref}

\begin{document}


\title{Invisible Higgs from forward muons at a muon collider}

\author{Maximilian Ruhdorfer}
 \email{m.ruhdorfer@cornell.edu}
\affiliation{%
 Laboratory for Elementary Particle Physics, Cornell University, Ithaca, NY 14853, USA
}%


\author{Ennio Salvioni}
 \email{ennio.salvioni@unipd.it}
\affiliation{
 Dipartimento di Fisica e Astronomia, Universit\`a di Padova and \\ INFN, Sezione di Padova, Via Marzolo 8, 35131 Padua, Italy
}%
\author{Andrea Wulzer}
\email{andrea.wulzer@cern.ch}
\affiliation{%
{{Institut de F\'{\i}sica d'Altes Energies (IFAE), The Barcelona Institute of Science and Technology (BIST),
Campus UAB, 08193 Bellaterra, Barcelona, Spain}} and \\
{{ICREA, Instituci\'o Catalana de Recerca i Estudis Avan\c{c}ats, 
Passeig de Llu\'{\i}s Companys 23, 
08010 Barcelona, Spain}}
}%



\begin{abstract}
We propose to probe the Higgs boson decay to invisible particles at a muon collider by 
observing the forward muons that are produced in association with the Higgs in the $Z$-boson fusion channel. An excellent sensitivity is possible in line of principle, owing to the large number of produced Higgs bosons, provided a forward muon detector is installed. We find that the resolution on the measurement of the muon energy and angle will be the main factor limiting the actual sensitivity. This poses tight requirements on the forward muon detector design.
\end{abstract}

\maketitle


\section{\label{sec:intro}Introduction}

The possibility of building a muon collider with centre of mass energy of 10~TeV or more and with high luminosity~\cite{Delahaye:2019omf} has received increasing attention in the last few years and is being actively pursued (see~\cite{Accettura:2023ked} for a review) by the International Muon Collider Collaboration (IMCC). Such collider would offer innumerable and varied physics opportunities, ranging from the direct access to the 10~TeV energy scale to the availability of a large effective luminosity for vector boson collisions at the scale of 1~TeV or below. The physics potential of the muon collider as a ``vector boson collider''~\cite{Han:2020uid} has been outlined in~\cite{Buttazzo:2018qqp,Ruhdorfer:2019utl,Liu:2021jyc} for the search of new particles produced in the Vector Boson Fusion (VBF) process, for the search for new phenomena in Standard Model (SM) scatterings initiated by vector bosons (VBS processes)~\cite{Costantini:2020stv,Han:2020pif,Buttazzo:2020uzc} and for precise measurements of the single Higgs couplings~\cite{Han:2020pif,Forslund:2022xjq}.

\begin{figure}[t]
\includegraphics[width=0.35\textwidth]{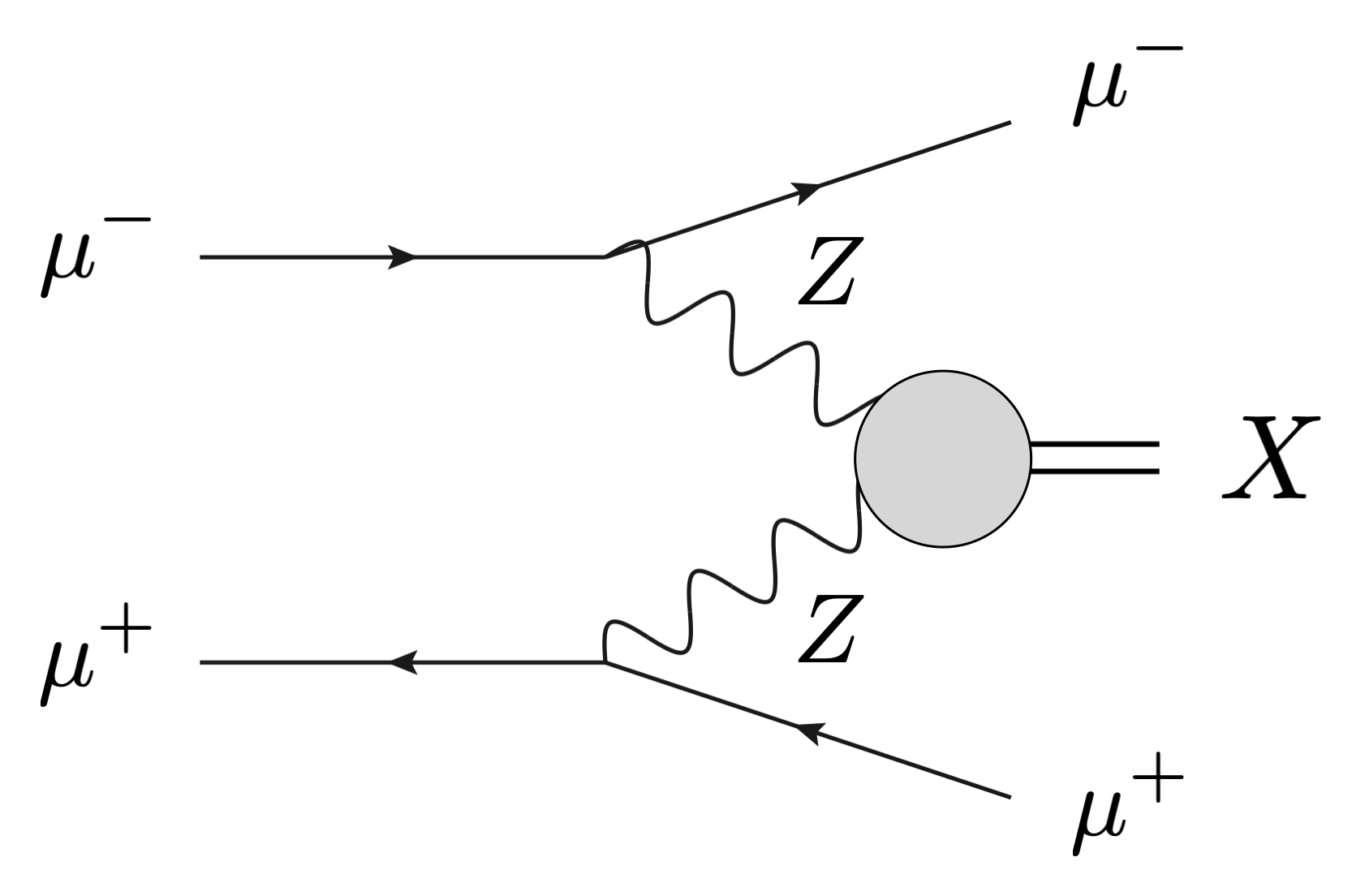}
\caption{\label{fig:vbf} Schematics of an effective $Z$ bosons collision producing a generic final state $X$. $Z$-fusion Higgs boson production, $X=h$, is the main focus of the present paper.}
\end{figure}

The VBF or VBS processes are schematically represented in Fig.~\ref{fig:vbf}. They proceed through the collinear emission of nearly on-shell vector bosons from the incoming muons. The vector bosons collide producing some final state ``$X$'' such as the Higgs boson, in the process considered in the present work. The on-shell fermion and anti-fermion emerge from the splitting as real final-state particles. If the emitted vector bosons are charged $W$ bosons, the initial muons are turned into invisible neutrinos. The emission of neutral bosons such as the $Z$ or the photon are instead accompanied by potentially detectable final-state muons, offering novel handles for the observation and the study of VBF and VBS processes. 

The kinematics of the process is conveniently described in the effective vector boson approximation~\cite{Kane:1984bb,Dawson:1984gx,Chanowitz:1985hj,Kunszt:1987tk,Borel:2012by} by factorising the emission of the vector bosons into universal splitting functions that are independent of the nature of the subsequent scattering process.
The typical transverse momentum of the effective $Z$ boson---and in turn the one of the final muon--- is around the mass of the boson, $p_\bot\sim m_Z$. The $p_\bot$ spectrum is almost entirely above one tenth of $m_Z$. 

The energy of the emitted bosons depends on the invariant mass of the $X$ system. If the invariant mass is of hundreds of GeV or less (e.g., $m_X=m_h$ in the case of Higgs production), the energy of the $Z$ is a small fraction of the initial muon energy. Therefore the final state muon carries away almost all of the beam energy $E_{\rm{b}}=5$~TeV and thus for $p_\bot\sim m_Z$ it has a small typical angle $\theta\sim p_\bot/E_{\rm{b}}=18~{\rm{mrad}}$ from the beam line. The invariant mass $m_X$ is larger than hundreds of~GeV if $X$ is a heavy new physics particle or if $X$ consists of a pair of SM particles and we apply an invariant mass cut in order to study their interaction with the $Z$ at the 1~TeV scale. In this case, the energy of the final muons is below the beam energy, the muons are less forward and a priori easier to detect. An angular coverage for muon detection up to around a pseudo-rapidity $|\eta|<7$, i.e.~$\theta \gtrsim 0.1\,m_Z/E_{\rm{b}}= 1.8~{\rm{mrad}}$, would definitely offer sensitivity to the entire $p_\bot$ spectrum of the forward muons associated with the emission of effective $Z$ bosons in all VBF or VBS processes of interest at the muon collider. 

An extended angular coverage above pseudo-rapidity 7 could be of interest to study the emission of effective photons rather than $Z$ bosons, because the effective photon $p_\bot$ distribution extends much below the 100~GeV scale, down to the muon mass. However, since the distribution is logarithmic, good sensitivity to the effective photon emission is expected even with angular coverage $|\eta|<7$. The observation of forward muons with rapidity up to around 7 or 6 thus signals the occurrence of a generic ``neutral'' VBF or VBS process where the colliding bosons could be either $Z$-bosons or photons, a priori. In the case of single-Higgs production, photons do not play any role because their coupling to the Higgs is small.

The current design of the muon collider machine-detector interface foresees the installation of two conical absorbers along the beam line---with the tips pointing towards the interaction point---in order to shield the detector from the radiation induced by the decay of the colliding muons. The absorbers limit the angular coverage of the main detector to $\theta > 10^{\rm o}=175~{\rm{mrad}}$~\cite{Accettura:2023ked}. This corresponds to $|\eta| < 2.44$, which is not sufficient to detect the forward muons produced in neutral VBF and VBS events. Fortunately, TeV-energy muons are penetrating particles that cross the absorbers and possibly other elements of the collider. Unlike all other species of particles, whose detection is possible only in the central region, muons could thus in principle be detected also in the pseudo-rapidity range from $2.5$ to $7$ if a dedicated forward muon detector was installed. 

The IMCC plans~\cite{Accettura:2023ked} to study the design of a forward muon detector, and in fact the possibility of detecting forward muons has long been included in the muon collider \texttt{DELPHES} card~\cite{deFavereau:2013fsa,delphes_card_mucol}. However, the assessment of the feasibility and of the performances of such detector has just started. Studying the physics potential of the forward muon detector provides useful guidance to the IMCC design study and informs on the requirements that would be needed in order to achieve specific goals.

The need for a forward muon detector first emerged~\cite{Ruhdorfer:2019utl} in the study of Higgs portal models. These are important new physics targets for future colliders aimed at probing the Higgs sector from multiple angles, including its possible connections with dark matter. The new physics particles coupled through Higgs portal interactions are copiously produced in VBF at the muon collider, but they must be stable and invisible in order to be a viable dark matter candidate. This motivates searches for invisible particles produced in the neutral VBF process. The forward muon detector will enable to tag this otherwise invisible signal and it will also allow to reject the background by exploiting the kinematics of the forward muons, provided the forward detector will be equipped to perform a measurement of the momentum and not just to identify the muons. 

A survey of several physics studies that exploit the forward muon detector, providing its physics case, is currently in preparation~\cite{toappear}. Among these studies, the novel analysis of the SM Higgs decay to invisibles that we propose in this paper poses the tightest requirements on the detector performances and in particular on the resolution of the muon energy and direction measurements.

The decay of the Higgs to invisible particles has not yet been established experimentally. LHC data result in an upper bound $\mathrm{BR}_{\rm inv} < 0.11$~\cite{ATLAS:2020kdi} on the invisible branching ratio at $95\%$~CL. A moderate improvement is expected at the High-Luminosity LHC (HL-LHC): $\mathrm{BR}_{\rm inv} < 0.028$~\cite{CMS:2017cwx,deBlas:2019rxi}, in the most optimistic scenario for systematic uncertainties. This sensitivity is very far from the prediction  $\mathrm{BR}_{\rm inv}^{\rm SM} = 1.2\cdot 10^{-3}$ from the SM decay $h \to ZZ^*\to 4\nu$. Among proposed future projects, electron-positron colliders such as FCC-ee and ILC will improve the sensitivity down to $\mathrm{BR}_{\rm inv} = 3\cdot 10^{-3}$~\cite{deBlas:2019rxi} (though a more optimistic study previously claimed $\sim1\cdot 10^{-3}$ reach~\cite{Chacko:2013lna}), but only the FCC-hh is expected to observe the SM invisible decay, with a projected sensitivity $\mathrm{BR}_{\rm inv} < 2.5\cdot 10^{-4}$~\cite{FCC:2019hfo} under the hypothesis of vanishing branching ratio. Clearly, after passing the SM threshold the relevant figure of merit becomes the expected sensitivity on the beyond-the-SM (BSM) invisible branching ratio $\mathrm{BR}_{\rm inv}^{\rm BSM}$, under the hypothesis of SM branching ratio. The BSM invisible branching ratio is due to the decay of the Higgs to new putative invisible particles, which may be stable or more generally long-lived, and is defined by the relation $\mathrm{BR}_{\rm inv}=\mathrm{BR}_{\rm inv}^{\rm SM}+ \mathrm{BR}_{\rm inv}^{\rm BSM}$. A variety of new physics scenarios foresee a sizeable $\mathrm{BR}_{\rm inv}^{\rm BSM}$, possibly even larger than the SM component~\cite{Curtin:2013fra,Cepeda:2021rql}. This provides a strong theoretical motivation for the study of invisible Higgs decays. In this paper we quantify the muon collider sensitivity accounting for all the relevant backgrounds and realistic beam parameters, as a function of the angular acceptance and resolution of the putative forward muon detector.

The paper is organised as follows. We start in Section~\ref{sec:truth} from an idealised setup where the incoming muons are perfectly monochromatic and parallel to the beam axis, and the final state muon angle and energy are measured perfectly. 
In Section~\ref{sec:beam} we include in the simulations the effect of the finite spread in energy and in angle of the colliding muon beams. The results obtained with these simulations provide the best attainable sensitivity with a forward detector that measures the momentum of the forward muons extremely precisely. Possibly realistic resolutions, included in Section~\ref{sec:results}, will emerge as the main factor limiting the sensitivity. We summarise our findings in Section~\ref{sec:conc}. The effect of variations of the assumed performances of the main detector, and sensitivity projections at a 3~TeV muon collider, are described in Appendix~\ref{app:main} and~\ref{app:3}, respectively.

\section{Truth-level distributions}\label{sec:truth}

In this section we study the signal $\mu^+ \mu^- \to \mu^+ \mu^- h$ with the Higgs decaying invisibly, and the relevant backgrounds, in an idealised setup where beam particles have exactly $E_{\rm{b}}=5$~TeV energy, and momentum along the beam axis. We also ignore the uncertainties in the measurement of the final muons. 

We consider events characterised by the detection of two opposite-charge muons, and a veto on any other object (photon, jet or lepton) within the coverage $|\eta|<2.44$ of the main detector. The energy or transverse momentum threshold above which the veto is effective will have to be estimated by a full simulation of the main detector, once available. We assume a common threshold of $p_\bot>$~20~GeV for all objects and study departures from this value in Appendix~\ref{app:main}. We consider final-state muons in opposite hemispheres (i.e., $\eta_{\mu^+}\cdot \eta_{\mu^-}<0$) with absolute pseudo-rapidity $|\eta_\mu|$ up to 7, and study in Section~\ref{sec:results} the impact of a reduced angular acceptance for the forward muon detector. We further require a lower energy threshold, $E_{\mu^\pm}>500$~GeV, for the muons to cross the absorbers and be detected. The contribution from virtual photons splitting to $\mu^+\mu^-$ (or from $Z$ bosons decay) is eliminated by a  loose angular separation cut $\Delta R_{\mu\mu}>0.4$. These selections define the ``baseline'' cuts for our analysis.

On top of the kinematic properties of the individual muons, signal and backgrounds are conveniently characterised and discriminated by the invariant mass $M_{\mu\mu}$, the azimuthal angular distance $\Delta \phi_{\mu\mu}$ and the total transverse momentum $P_\bot^{\mu\mu} =(p_{\mu^+} + p_{\mu^-})_\bot$ of the $\mu^+\mu^-$ pair. Other useful variables are the minimal muon energy, $E_{\mu}^{\rm min} = \mathrm{min}\,(E_{\mu^-}, E_{\mu^+})$, and the Missing Invariant Mass (MIM)
\begin{equation}\label{eq:MIM}
\mathrm{MIM} =  \sqrt{\left| (\Delta P)^2 \right|} \,,\;\; \Delta P = (2E_{\rm{b}}, \vec{0}\,) - p_{\mu^+} - p_{\mu^-}\,,
\end{equation} 
where $2E_{\rm{b}}=10$~TeV is the nominal centre of mass energy of the collider. $\Delta P$ is the difference between the total 4-momentum of the incoming muons with nominal energy and the total 4-momentum of the final muons. The absolute value in Eq.~\eqref{eq:MIM} ensures that MIM remains real and positive even in the presence of beam energy spread and experimental smearing of the final state muon momenta.

\begin{figure*}[t]
\includegraphics[width=\textwidth]{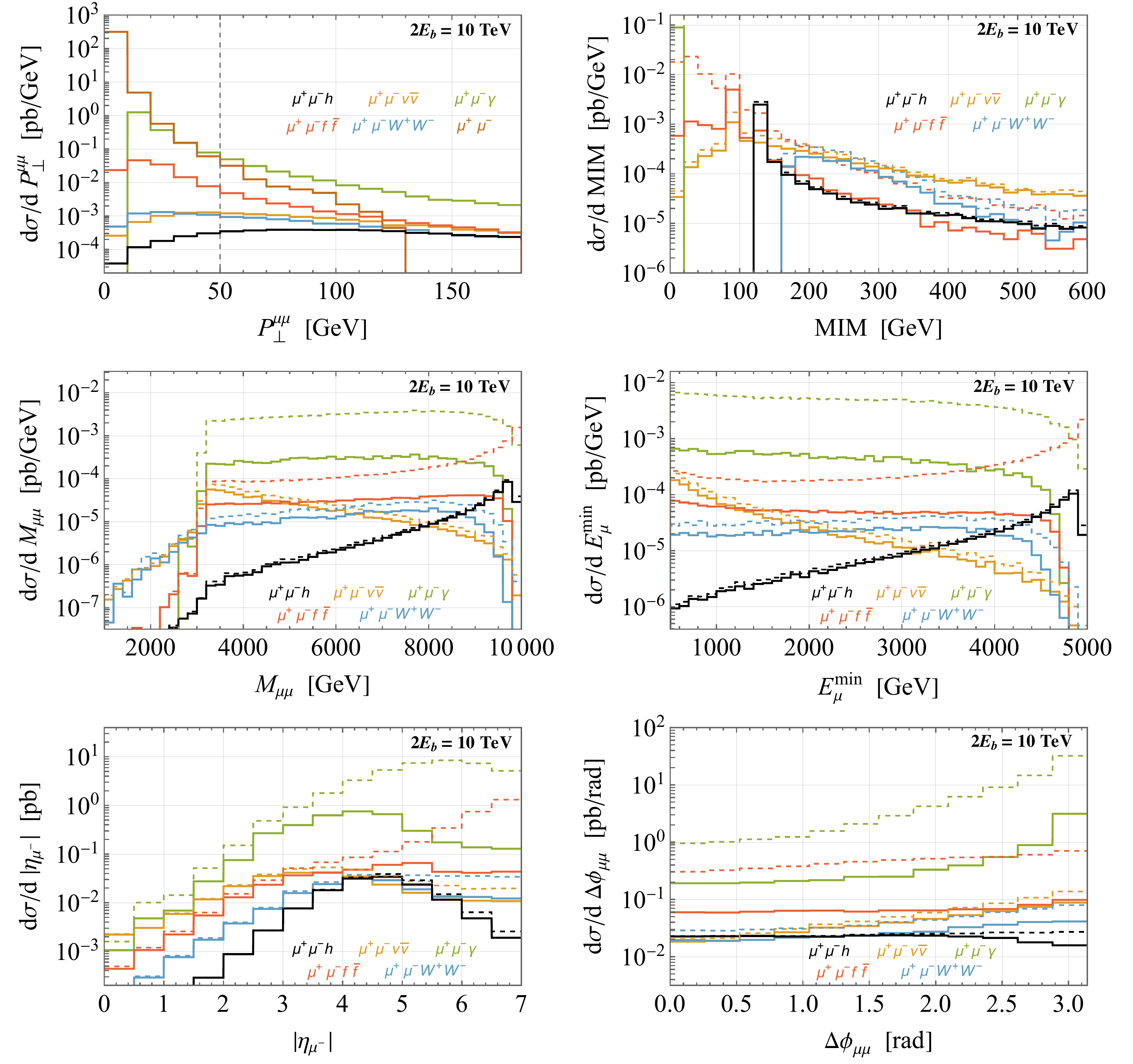}
\caption{\label{fig:truth} Key kinematic distributions at truth level. The signal cross section corresponds  to $\mathrm{BR}_{\rm inv} = 1$. In the top left panel, only baseline cuts are applied (i.e., $|\eta_\mu|<7$, $\eta_{\mu^+}\cdot \eta_{\mu^-}<0$, $\Delta R_{\mu\mu}>0.4$ and $E_{\mu^\pm}>500\,{\rm{GeV}}$, plus the veto with $p_\bot>20\,{\rm{GeV}}$ and $|\eta|<2.44$). In all the other panels, dashed lines correspond to baseline cuts, while solid lines also include $P_\bot^{\mu\mu} > 50$~GeV.}
\end{figure*}

Monte Carlo (MC) data samples for signal and backgrounds are generated using \texttt{MadGraph5\_aMC@NLO}~\cite{Alwall:2014hca}. Photon showering from \texttt{PYTHIA8}~\cite{Bierlich:2022pfr} is also performed on the signal and on some of the background samples (see below). Both final state radiation (FSR) and initial state radiation (ISR) showering are included. The backwards evolution needed for ISR requires the muon parton distribution function (PDF) of the muon to be employed in the fixed-order event generators. This is achieved in \texttt{MadGraph} by a simple modification of the electron PDF implementation~\cite{Frixione:2021zdp}. 
The relevant distributions for the signal are shown in black in Fig.~\ref{fig:truth}. The total cross-section is $62$~fb for unit branching ratio of the invisible decay of the Higgs, after the baseline cuts described above and $P_\bot^{\mu\mu} > 50$~GeV. The signal is characterised by a muon pseudo-rapidity between around $2.5$ and 7, in accordance with the estimates described in the Introduction. The muon $p_\bot$ is of order $m_Z$ as previously discussed, thus $P_\bot^{\mu\mu}$ is of order hundred GeV as the figure shows. Relevant backgrounds are those processes that can produce a forward and energetic $\mu^+\mu^-$ pair, but with some momentum unbalance in the transverse plane. They fall into two categories. 

The first class of backgrounds are those processes that produce invisible neutrinos, namely the final state $\mu^+ \mu^- \nu \bar{\nu}$ shown in orange in Fig.~\ref{fig:truth}. While we simulate this final state as a single process, we notice that it contains a number of different components. Subprocesses where the neutrinos are emitted from the decay of a $Z$ boson dominantly emerge from the radiation of the $Z$ from the elastic scattering $\mu^+ \mu^- \to \mu^+ \mu^-$. They are characterised by a resonant $Z$-pole peak in the MIM distribution, while the signal peaks at ${\rm{MIM}}=m_h$. However, we will see in the following section that the energy spread of the muon beams eliminates these narrow peaks.
The second component of the $\mu^+ \mu^- \nu \bar{\nu}$ background emerges from the $W$ boson produced in $\gamma W$ fusion, or radiated from the elastic muon scattering, and decaying as $W\to\mu\nu$. These processes produce a continuous spectrum in $M_{\mu\mu}$ and in MIM. A third component of the process comes from $Z$ bosons or low-virtuality photons emitted from the elastic process and decaying to muons. This component is however eliminated by the $\Delta R_{\mu\mu}$ and $E_{\mu^\pm}$ cuts.

The top left panel in Fig.~\ref{fig:truth} shows that the $P_\bot^{\mu\mu}$ distribution of the $\mu^+ \mu^- \nu \bar{\nu}$ background is very similar to the one of the signal. On the other hand, $P_\bot^{\mu\mu}$ is an important discriminant for other backgrounds, discussed below. For this reason, in Fig.~\ref{fig:truth} we show (with solid lines) the effect of a $P_\bot^{\mu\mu}>50$~GeV cut on the distributions. The cut has little impact on the $\mu^+ \mu^- \nu \bar{\nu}$ background. 

The second class of backgrounds are processes where the $\mu^+ \mu^-$ pair is produced in association with any type of object that cannot be vetoed because it falls outside the angular acceptance of the main detector or because it is softer than the assumed $20$~GeV $p_\bot$ threshold. 

The muons must be forward (while still below $|\eta_\mu| = 7$) for the process to be relevant. This naturally occurs for neutral VBF or VBS processes initiated by virtual $Z$ or photons. The largest such process is  $\mu^+\mu^-\rightarrow \mu^+\mu^- f\bar{f}$, where $f$ denotes any light quark or lepton with the exception of the muons, which are detected in the forward region. This final state includes VBF Higgs production with the Higgs decaying to $b\bar{b}$ or $\tau^+\tau^-$ as a subdominant contribution. It also includes the production of a virtual photon decaying to $f\bar{f}$. The corresponding singularity is eliminated by a 10~GeV cut on the invariant mass of the $f\bar{f}$ pair. The $\mu^+\mu^-\rightarrow \mu^+\mu^- \gamma$ process, discussed later in this section, accounts for the region below the cut. We also include the $\mu^+\mu^-\rightarrow \mu^+\mu^- W^+ W^-$ background, which is the largest vector boson or Higgs pair production process in neutral VBF~\cite{Costantini:2020stv}. This is estimated under the conservative assumption that all the $W$ bosons emitted outside the angular acceptance or with a $p_\bot$ lower than the threshold will pass the veto. Even with this setup, the $\mu^+\mu^- W^+ W^-$ background will be found to play a minor role for the sensitivity.

We have been unable to include photon showering in the $\mu^+\mu^- f\bar{f}$ process, which is thus simulated purely at tree-level and without muon PDFs. The first technical issue we encountered is that \texttt{MadGraph} event generation fails for this process when employing the muon PDFs, while PDFs are essential for ISR showering as previously mentioned. With PDFs, generation succeeds only with relatively strong lower $p_\bot$ cuts on all the final particles, in addition to the muon acceptance cut and the cut on the $f\bar{f}$ invariant mass. A second more conceptual difficulty concerns the choice of the showering scale in \texttt{PYTHIA8}. A large component of the process emerges from the $\gamma\gamma\to f\bar{f}$ fusion of low-virtuality effective photons. This is effectively an electromagnetic radiation process in itself, such that the adequate showering scale for photon radiation from the muon legs should be commensurate to the virtuality of the splitting rather than to the hardness of the final fermions. A sophisticated showering scheme should be adopted, which however does not seem to be easily available in \texttt{PYTHIA8}. We consider that this limitation of the simulation will not harm the accuracy of our results, because the effects of showering are minor in general and because $\mu^+\mu^- f\bar{f}$ will turn out not to be the dominant background.

Elastic scattering $\mu^+\mu^-\rightarrow \mu^+\mu^-$ also produces forward muons due to the $t$-channel enhancement. The emission of real photons that are either too soft or too forward to be detected generates a significant amount of MIM and $P_\bot^{\mu\mu}$. Ideally, one would like to simulate this process by generating a merged sample of $\mu^+ \mu^- \to \mu^+\mu^-$ plus additional photons and match it with QED showering. However, this is not straightforward to achieve with any of the currently available multi-purpose MC generators, in contrast with the case of QCD radiation. We then proceed as follows. We first generate a sample without extra photons and shower it with \texttt{PYTHIA8}.\footnote{The \texttt{PYTHIA8} settings must be modified to enforce the $t$ Mandelstam variable as the cutoff of the shower. We verified that our results agree with the native \texttt{PYTHIA8} $\mu^+\mu^-$ process results.} This produces the $P_\bot^{\mu\mu}$ distribution shown in brown color in the top left panel of Fig.~\ref{fig:truth}. The distribution extends above around $50$~GeV, where the signal starts, but with a very weakly populated tail that makes event generation above the analysis cut $P_\bot^{\mu\mu}>50$~GeV cumbersome. Furthermore, showering is arguably not the adequate description of the process in the large $P_\bot^{\mu\mu}$ tail. We thus simulate the tree-level process $\mu^+ \mu^- \to \mu^+\mu^-\gamma$ with a lower cut of $10$~GeV on the photon $p_\bot$ and on the photon-muon invariant mass in order to avoid the showering region. The resulting $\mu^+\mu^-\gamma$ simulation reproduces the results of showering quite well for $P_\bot^{\mu\mu}\approx 50$~GeV, but with a bigger tail for large $P_\bot^{\mu\mu}$. We employ the former simulation for the description of the elastic scattering background and we do not include the $ \mu^+\mu^-$ showered sample to avoid double-counting.\footnote{Including this sample as an additional background is found not to affect our results, because it is subdominant to $\mu^+\mu^-\gamma$.}

The most peculiar feature of the $\mu^+\mu^-\gamma$ background is the sharp peak at ${\rm{MIM}}=m_\gamma=0$. This peak is unphysical because it would be smeared out by the radiation of extra photons, which is not present in our purely tree-level simulation. However, the smearing due to showering is subdominant to the one due to the finite energy spread of the incoming muons, to be discussed in the next section. The tree-level modelling of the ${\rm{MIM}}$ distribution is thus adequate, as we also verified explicitly by comparing with showered event samples. We also generated the process with one extra matrix-element photon emission, $\mu^+\mu^-\gamma\gamma$, and checked that its addition does not affect the distributions after beam effects are included.

The \texttt{WHIZARD}~\cite{Moretti:2001zz,Kilian:2007gr} package has been employed to validate the distributions of the signal and of some of the backgrounds. An extensive and detailed comparison between \texttt{MadGraph5\_aMC@NLO} and \texttt{WHIZARD} predictions will be presented elsewhere~\cite{toappear}. 

\section{Beam effects}\label{sec:beam}

The truth-level MIM distributions in Fig.~\ref{fig:truth} offer in principle a very good handle to discriminate the signal from the backgrounds. However, the characteristic MIM shapes of the signal and of the backgrounds at around 100~GeV are highly sensitive to any imperfections in the knowledge of the final and of the initial muon momenta, because of cancellations, as we will readily see. It is thus mandatory to include in the simulations a realistic treatment of the muon beams, accounting for the finite beam energy spread (BES) and beam angular spread (BAS) at the interaction point. At a high-energy muon collider, their size is expected to be $\delta E/E  =\delta_{\rm BES}=0.1\% $ and $\delta \theta  = 0.6$~mrad, respectively \cite{Accettura:2023ked}. The uncertainties on the measurement of the momentum of the final-state muons  will we included in the next section.

The BES and the BAS cause a departure from the nominal total momentum of the initial state, $(2E_{\rm{b}},\vec{0}\,)$, which in turn impacts the kinematics of the outgoing muons. Clearly the MIM is still calculated according to Eq.~\eqref{eq:MIM}, since the initial momentum is not known on an event-by-event basis. The beam smearing can be regarded as an uncertainty on the knowledge of the true initial muon momenta, and in turn of the true $\Delta{P}$. The energy and longitudinal components of $\Delta{P}$ result from a cancellation between the initial and final muons momenta, which are of order $5$~TeV. A relatively small MIM of order 100~GeV is thus strongly affected even by a small relative spread of the muon beams.

We begin with a discussion of the BES, which has two main effects. First, the centre of mass energy of the initial muons, $\sqrt{s}$, differs from the nominal collider centre of mass energy~$2E_{\rm{b}}=10$~TeV. Second, the centre of mass frame of the muon collision does not coincide with the detector frame, the two being related by a Lorentz boost along the beam axis. The BES simulation is not implemented in \texttt{MadGraph}, therefore we account for it by proceeding as follows. 

We generate truth-level $\mu^+\mu^-$ collision events in the centre of mass frame, for different values of the centre of mass energy $\sqrt{s}$. If the energy of each beam is Gaussian distributed, $\sqrt{s}$ is also approximately Gaussian, with mean $2E_{\rm{b}}$ and standard deviation $\sigma=\sqrt{2}\, \delta_{\rm BES}E_{\rm{b}}$. We sample this distribution at three fixed values of $\sqrt{s}$ given by $\{2E_{\rm{b}} -\sigma, 2E_{\rm{b}}, 2E_{\rm{b}} +\sigma\}$, obtaining three event datasets that we combine with equal weights of $1/3$. Next, we introduce the boost of the centre of mass frame. The boost distribution conditional to $\sqrt{s}$ is approximately Gaussian, with zero mean and with standard deviation $\sigma/(2\sqrt{s})\simeq\, \delta_{\rm BES}/(2\sqrt{2})$. For each truth-level event we generate the boost by sampling from this distribution and we apply the corresponding Lorentz transformation to the final-state particles. We have checked that this simple method to simulate the BES is in good agreement with the BES implementation available in \texttt{WHIZARD}.

\begin{figure}[t]
\includegraphics[width=0.46\textwidth]{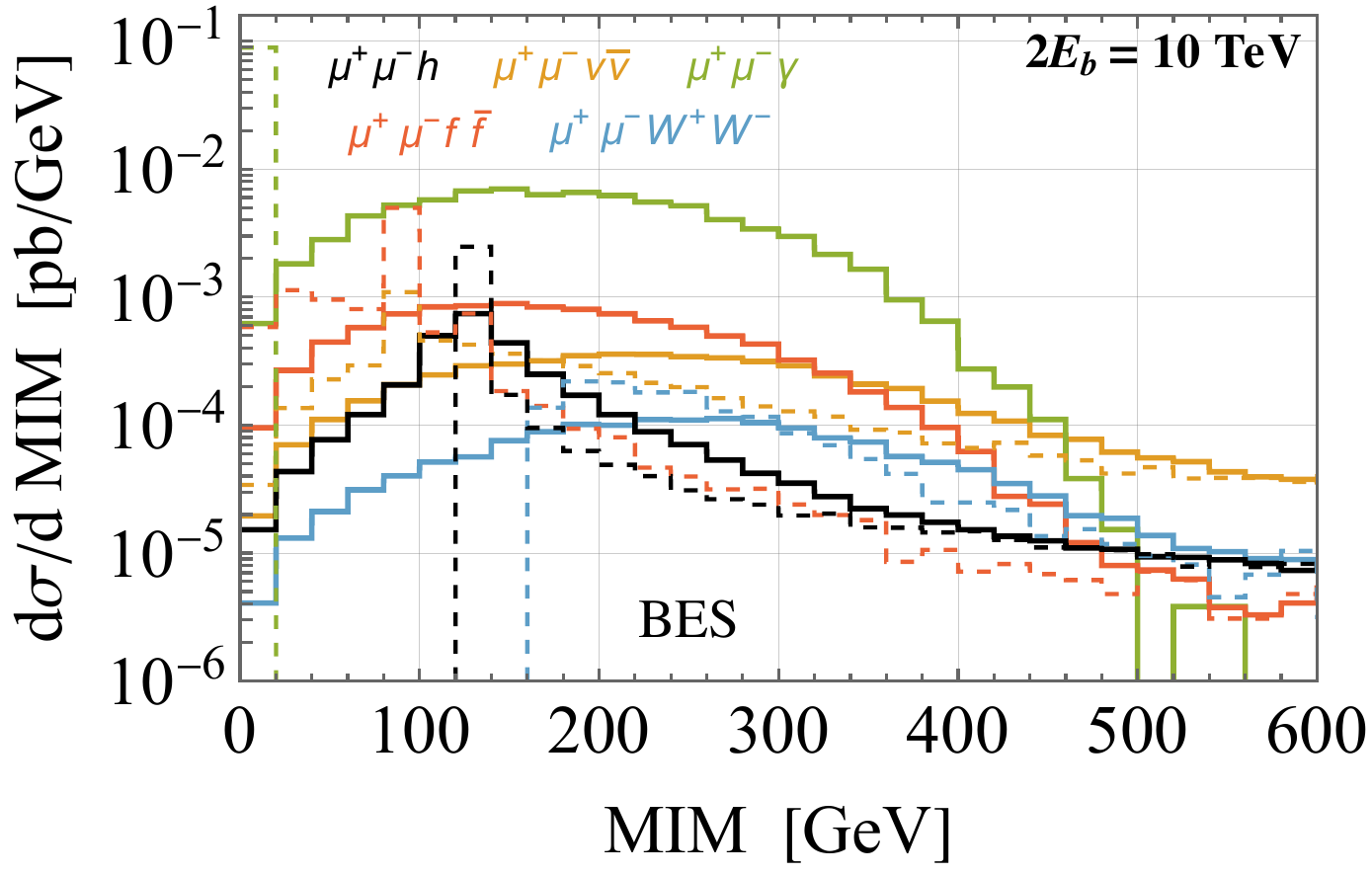}
\caption{\label{fig:BES} MIM distributions after the inclusion of beam energy spread (solid) compared to truth level (dashed). The cut $P_\bot^{\mu\mu} > 50$~GeV has been applied.}
\end{figure}

The BES has a minor impact on all distributions, which remain essentially identical to the truth-level ones in Fig.~\ref{fig:truth}, with the exception of the MIM distribution, as expected. This is shown in Fig.~\ref{fig:BES}. For the signal, and the $\mu^+\mu^-\nu\bar\nu$ or $\mu^+\mu^-f\bar{f}$ backgrounds, the effect is a considerable broadening of the distribution that eliminates the sharp resonant peaks at, respectively, $m_h$ and $m_Z$. The effect is less strong on $\mu^+\mu^-W^+W^-$, as the truth-level distribution is already rather broad. The major BES effect in this case is to populate the $\textrm{MIM}<2\,m_W$ region.

In the case of the $\mu^+\mu^-\gamma$ background, instead, the BES effect is dramatic: the distribution is turned from a sharp peak at $\mathrm{MIM} = 0$ into a wide plateau with a maximum at $\mathrm{MIM}\sim 150$~GeV. This can be understood by exploiting momentum conservation and writing Eq.~\eqref{eq:MIM} in the form
\begin{equation}\label{eq:MIM1}
\mathrm{MIM}^2 = \Big| \big( (2E_{\rm{b}}, \vec{0}\,) + p_\gamma - p_{\mu^-}^{\rm in} - p_{\mu^+}^{\rm in} \big)^2 \Big|\,,
\end{equation}
where $p_{\mu^\pm}^{\rm in}$ are the actual momenta of the initial-state muons and $p_\gamma$ is the momentum of the undetected photon. Clearly the MIM would vanish if the total momentum of the initial muons was equal to $(2E_{\rm{b}}, \vec{0}\,)$, because $p_\gamma^2=0$. However, in the presence of the BES, at the $1\sigma$ level we encounter configurations such as 
\begin{equation}
p_{\mu^-}^{\rm in} + p_{\mu^+}^{\rm in}= E_{\rm{b}}\left(2+\delta_{\rm BES}, \vec{0}_\bot,\delta_{\rm BES}\right)\,.
\end{equation}
By substituting in Eq.~\eqref{eq:MIM1}, we obtain the estimate
\begin{equation}
\mathrm{MIM}^2 \sim (150\;\mathrm{GeV})^2 \,\frac{\delta_{\rm BES}}{10^{-3}}\, \frac{|p^\gamma_z|}{0.2E_{\rm{b}}} \left( \frac{2E_{\rm{b}}}{10\;\mathrm{TeV}} \right)^2,
\end{equation}
where $|p^\gamma_z| \gg p_\bot^\gamma$ was assumed, since the photon is typically emitted in the forward direction. The emission of photons with $|p^\gamma_z| \sim 0.2E_{\rm{b}}\sim1$~TeV is relatively likely, owing to the collinear enhancement of the photon splitting. In turn, these effects are responsible for the change of shape of the MIM distribution for the $\mu^+\mu^-\gamma$ background, observed in Fig.~\ref{fig:BES}. These considerations also show that relatively large values of the MIM can be attained in the $\mu^+\mu^-\gamma$ background only in events characterised by the emission of a rather energetic photon. In these events, the energy of either the final muon or anti-muon is significantly smaller than $E_{\mathrm{b}}$. Therefore, we will still be able to partially eliminate them by a cut on $E_\mu^{\rm min}$.

The starting point for the BAS simulation is again truth-level samples generated in the centre of mass frame of the initial muons and with center of mass energy $2E_{\rm{b}}$. We assume that the polar angle of each beam muon is Gaussian distributed around zero, with standard deviation $\delta \theta  = 0.6$~mrad, while the azimuthal angle is uniformly distributed. For each event, we determine the direction of each muon by throwing the angles from these distributions. We take the two muons to have the same energy, which is computed by imposing that the centre of mass energy is equal to $2E_{\rm{b}}$, and we construct the 4-momenta of the initial muons. Then we consider the Lorentz transform that brings the initial muons in their centre of mass frame, and apply its inverse to all the particles in the event. 

\begin{figure}[t]
\includegraphics[width=0.46\textwidth]{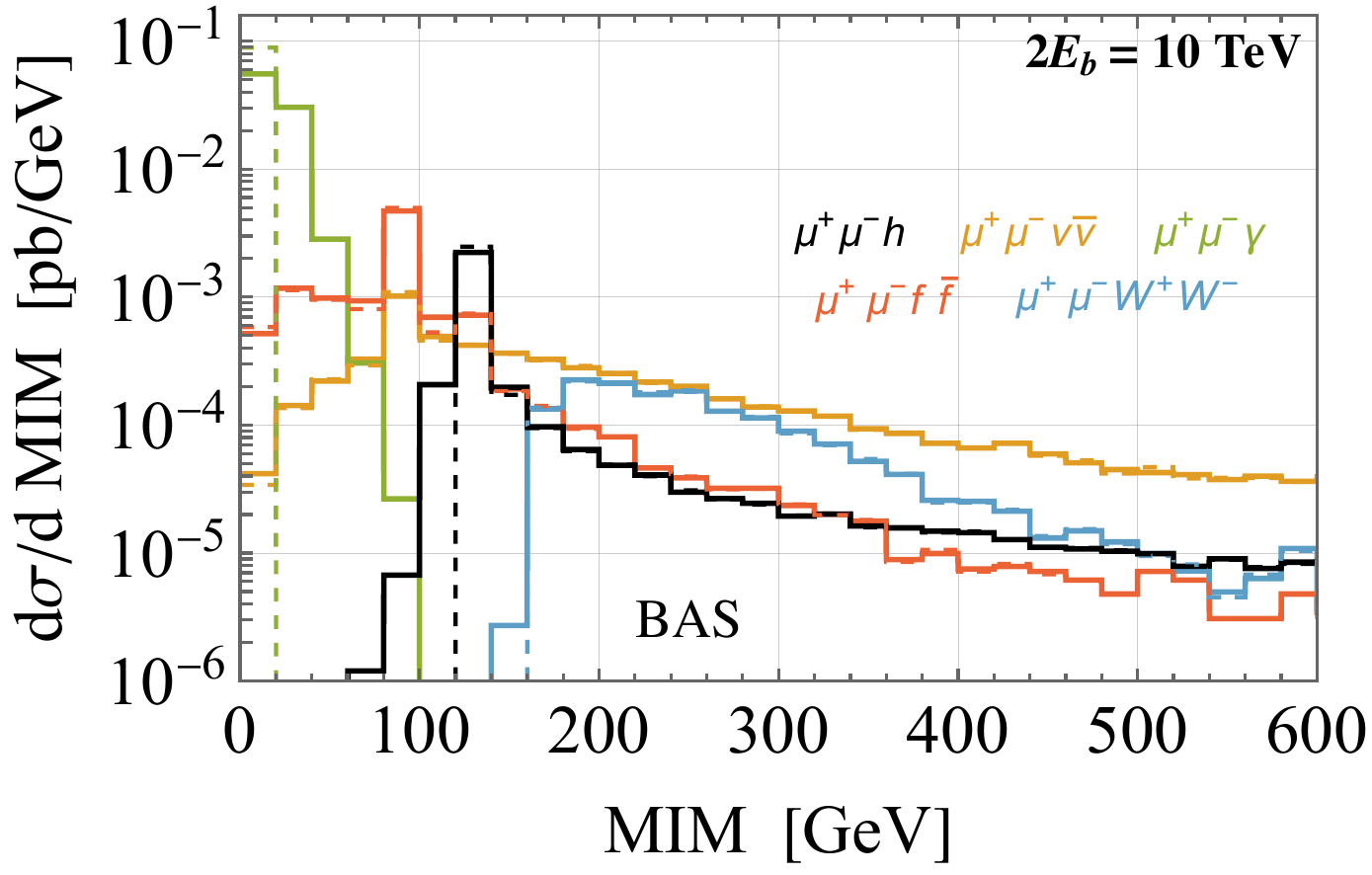}
\caption{\label{fig:BAS} MIM distributions after the inclusion of beam angular spread (solid) compared to truth level (dashed). The cut $P_\bot^{\mu\mu} > 50$~GeV has been applied.}
\end{figure}

The effect of the BAS is found to be minor for both signal and backgrounds on all kinematic variables. The largest effect, shown in Fig.~\ref{fig:BAS}, is a smearing of the MIM for the $\mu^+\mu^-\gamma$ background to nonzero values, but this is anyway much smaller than the impact of BES shown in Fig.~\ref{fig:BES}. Therefore, the BAS is neglected in the following.

\section{Results}\label{sec:results}

The target luminosity of the 10~TeV muon collider~\cite{Accettura:2023ked}, of $10\,{\rm{ab}}^{-1}$, will produce 10 million Higgs bosons in total, mostly in the charged vector boson fusion process. Around one million of them---620'000, considering the cross-section of $62$~fb after the $P_\bot^{\mu\mu}>50$~GeV cut---are produced in the $Z$ boson fusion process. With the SM branching ratio of $1.2\cdot 10^{-3}$, about 1'000 invisible decays will be available, allowing in principle not only to observe the SM invisible decay, but also to measure $\mathrm{BR}_{\rm inv}$ with few percent relative accuracy. A BSM contribution $\mathrm{BR}_{\rm inv}^{\rm BSM}$ could thus be probed at the $10^{-4}$ level.

\begin{figure}[t]
\includegraphics[width=0.48\textwidth]{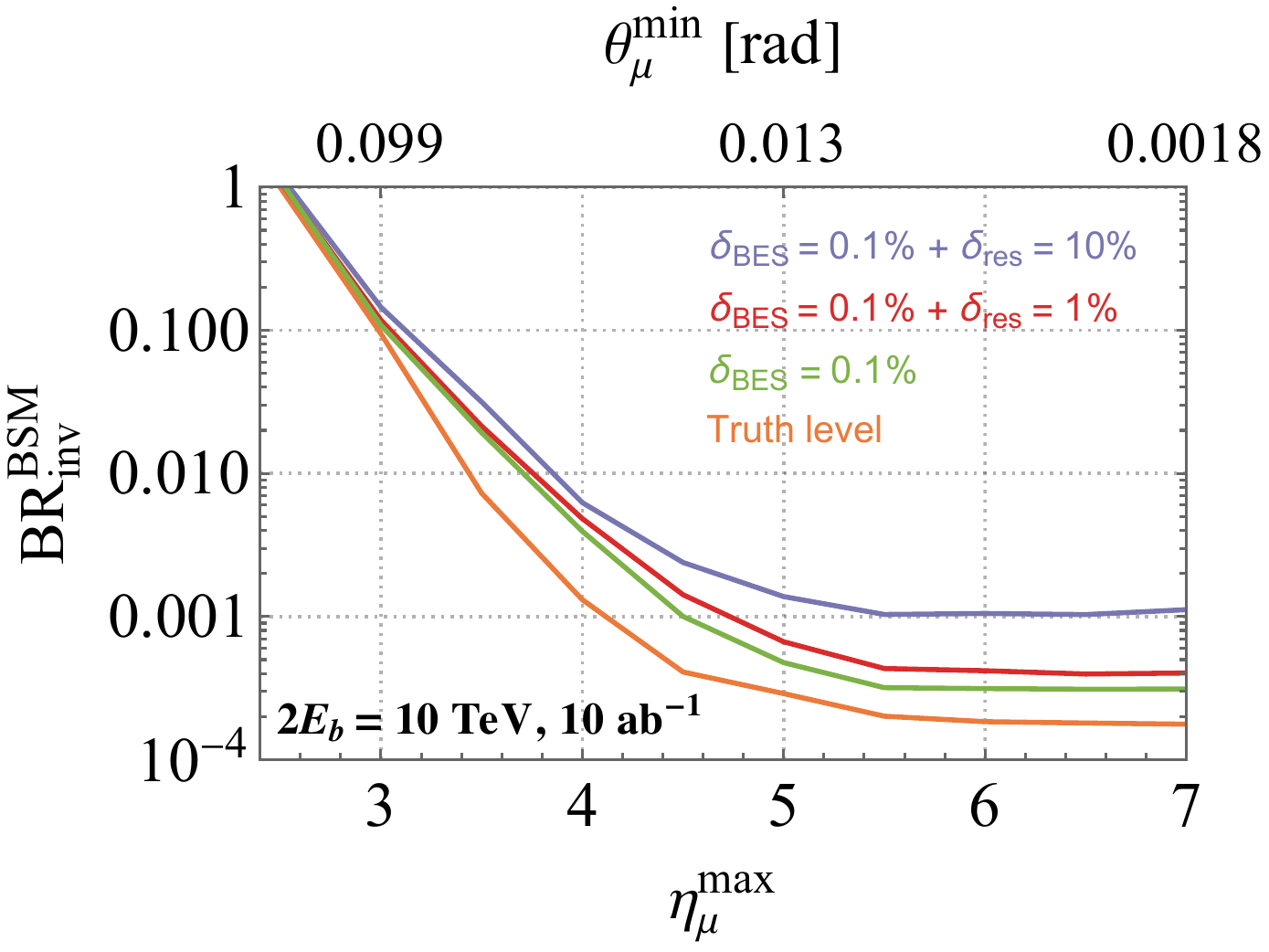}
\caption{\label{fig:BR10TeV} 95\% CL limit on BR$_{\rm inv}^{\rm BSM}$ at a 10~TeV muon collider with $10$~ab$^{-1}$, as a function of the angular acceptance $2.44 < |\eta_\mu| < \eta_\mu^{\rm max}$ of a forward muon detector.}
\end{figure}

Attaining $10^{-4}$ level sensitivity would require strongly discriminant kinematical features enabling the design analysis cuts that eliminate the background while preserving a large fraction of the signal. This is indeed possible with the truth-level distributions in Fig.~\ref{fig:truth}, very discriminating variables being the MIM, $E_\mu^{\rm{min}}$ and $M_{\mu\mu}$. Thanks to the excellent background rejection, the sensitivity to BR$_{\rm inv}^{\rm BSM}$ would be in fact around $10^{-4}$ for truth-level events, as Fig.~\ref{fig:BR10TeV} shows. A forward muon detector coverage up to $\eta_\mu^{\rm{max}}=6$ would be sufficient to achieve this sensitivity. 

Interestingly enough, nearly perfect selection cuts can be designed also in the presence of a realistic spread of the beam energy. The BES reduces the discriminating power of the MIM distribution, as shown in Fig.~\ref{fig:BES}, but stronger lower cuts on $E_\mu^{\rm{min}}$ and $M_{\mu\mu}$ can still eliminate the background with limited signal rejection. The BR$_{\rm inv}^{\rm BSM}$ sensitivity (see again Fig.~\ref{fig:BR10TeV}) thus remains at around $10^{-4}$ even in the presence of beam effects. We conclude that if the muon collider beam energy spread will be at the level of $10^{-3}$ as foreseen, its effect will not limit the sensitivity to the Higgs invisible decay significantly. We saw in the previous section that the beam angular spread does not play an important role.

The finite resolution in the measurement of the energy and of the angle of the final muons are instead expected to play a major role in limiting the sensitivity. We include these effects in our simulations by proceeding as follows. 

We assume a constant relative uncertainty $\delta_{\rm res}$ on the muon energy measurement, which we simulate by throwing the measured energy of each muon from a Gaussian distribution centred around the true energy $E_\mu$ and with standard deviation $\delta_{\rm res}\cdot E_\mu$. The actual response function of the forward muon detector will most likely not be centred at $E_\mu\,$, because the muons will lose energy while crossing the absorbers, nor will it be symmetric around the centre. However, preliminary results obtained with a more realistic response function confirm the adequacy of employing a Gaussian smearing.\footnote{We thank Daniele Calzolari and Federico Meloni for sharing their initial estimate of the response function with us.} The smearing is applied only to muons outside the acceptance of the main detector, $|\eta_\mu| > 2.44$. The muons inside the acceptance of the main detector play a very limited or no role in the analysis, and moreover we expect that their energy will be measured more accurately. 

In order to simulate the uncertainty in the measurement of the muon direction we generate a polar angle $\theta$, thrown from a Gaussian centred at zero and with standard deviation $\Delta\theta$, and an azimuthal angle $\phi$ with uniform distribution. The measured muon direction is chosen to form an angle of $\theta$ with the true direction $\hat{n}_{\rm{tr}}$. The orientation of the measured direction in the plane transverse to $\hat{n}_{\rm{tr}}$ is determined by the azimuthal angle $\phi$.

\begin{figure*}[t!]
\includegraphics[width=\textwidth]{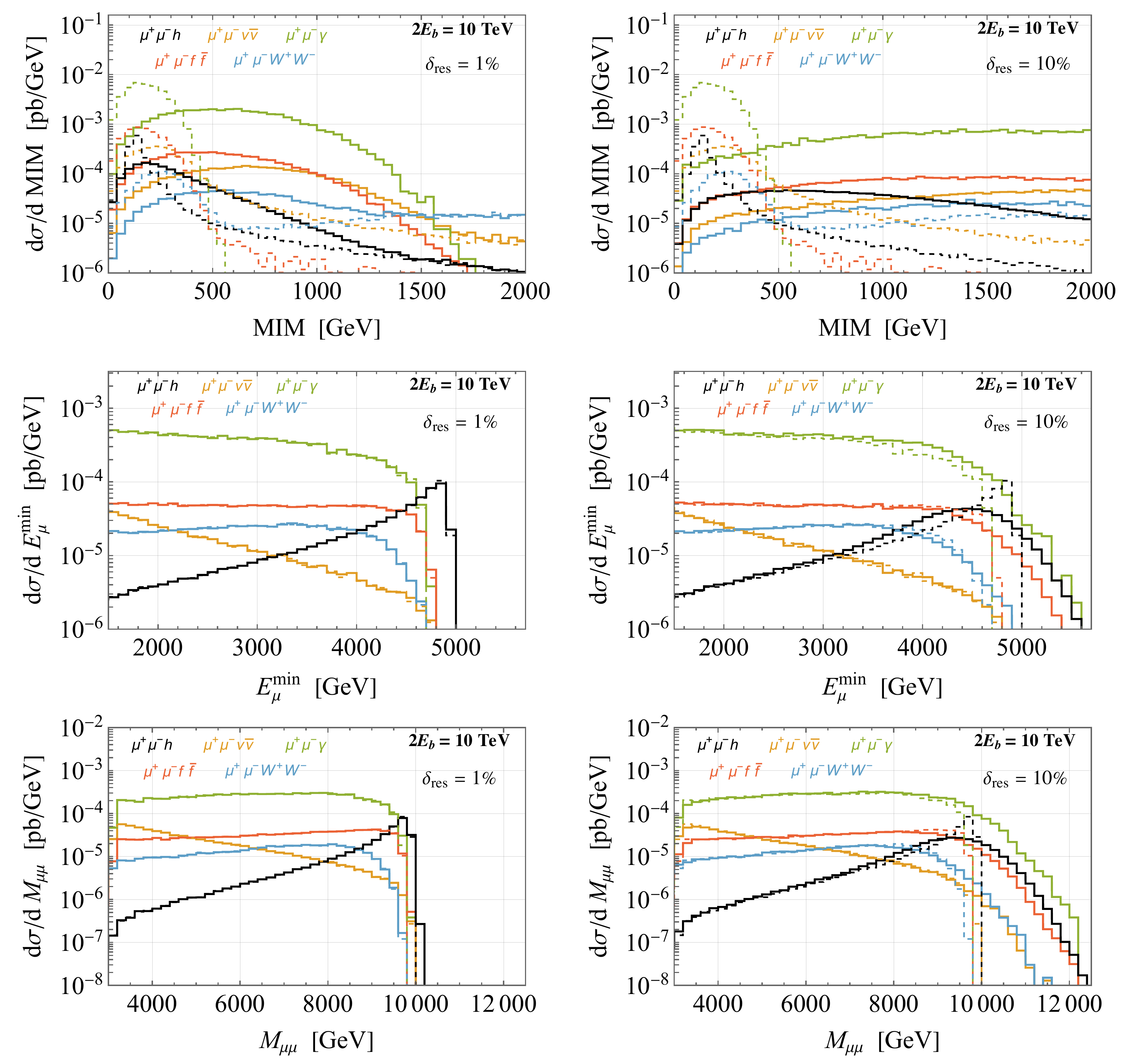}
\caption{\label{fig:smeared} Key kinematic distributions after BES and smearing of the forward muon energies (solid), compared to those including only BES (dashed). The left~(right) panels correspond to a forward detector resolution $\delta_{\rm res} = 1\,(10)\%$. The signal cross section corresponds to $\mathrm{BR}_{\rm inv} = 1$. In all panels, the baseline cuts and $P_\bot^{\mu\mu} > 50$~GeV are applied.}
\end{figure*}

The most effective distributions for signal/background discrimination, namely $\mathrm{MIM}, E_\mu^{\rm min}$ and $M_{\mu\mu}$, are significantly affected by the energy measurement uncertainty, as Fig.~\ref{fig:smeared} shows. All other distributions remain essentially identical to the truth-level ones in Fig.~\ref{fig:truth}. 

The effect on the MIM is considerable even with  energy resolution as small as $\delta_{\rm res} = 1\%$. However, although broadened, the MIM distribution for the signal maintains a peak at around $200$~GeV that is still useful to reject the background. Furthermore, the $E_\mu^{\rm min}$ and $M_{\mu\mu}$ distributions are almost unaffected by the $\delta_{\rm res} = 1\%$ smearing and retain a good discriminating power. Effective analysis cuts can thus be designed and the sensitivity degradation due to the inclusion of the energy uncertainty is marginal as we see in Fig.~\ref{fig:BR10TeV} by comparing the red and the green curves. An optimised cut-flow for an angular acceptance of $\eta_\mu^{\rm max} = 6$ is reported in  Table~\ref{tab:cut_flow_1p}.  After the cuts, 130 invisible decays are expected for SM branching ratio, with a background of around 600. The SM decay could thus be easily observed and, in the case of agreement with the SM, BSM effects could be bounded to  $\mathrm{BR}_{\rm inv}^{\rm BSM} < 4 \cdot 10^{-4}$ at $95\%$ CL. At the exclusion, the ratio of signal to background is relatively large, $\mathrm{S/B} \approx 6\%$. Background estimates should be possible with better or comparable accuracy. Systematic uncertainties are thus not expected to reduce the sensitivity strongly.

\renewcommand{\tabcolsep}{6pt}
\renewcommand{\arraystretch}{1.4}
\begin{table*}[t]
\centering
\begin{tabular}{ lcccccc }    
\hfil [number of events, $10$~ab$^{-1}$] & BSM signal &  $\mu^+\mu^-\bar{\nu}\nu$ &  $\mu^+\mu^- \gamma$ & $\mu^+\mu^- \bar{f} f$ & $\mu^+\mu^- W^+ W^-$  & $\mu^+\mu^- (h\rightarrow \text{inv})_{\rm SM}$\\\hline
baseline $\&\, P_{\bot}^{\mu\mu} > 50$~GeV  & $6.2\cdot 10^{5}\cdot{{\rm{BR}}_{\rm inv}^{\rm BSM}}$ & $1.1\cdot 10^{6}$  & $1.3\cdot 10^{7}$ & $1.3\cdot 10^{6}$ &  $6.2\cdot 10^{5}$  & $7.4\cdot 10^{2}$\\
$\text{MIM} < 0.8$~TeV & $5.6\cdot 10^{5}\cdot{{\rm{BR}}_{\rm inv}^{\rm BSM}}$ & $6.3\cdot 10^{5}$   & $1.0\cdot 10^{7}$ & $9.4\cdot 10^{5}$ & $1.8\cdot 10^{5}$  & $6.7\cdot 10^{2}$\\
$|\Delta \eta_{\mu\mu}| > 8$ & $4.8\cdot 10^{5}\cdot{{\rm{BR}}_{\rm inv}^{\rm BSM}}$ & $2.3\cdot 10^{5}$   & $5.8\cdot 10^{6}$ & $6.3\cdot 10^{5}$ & $1.3\cdot 10^{5}$  & $5.8\cdot 10^{2}$\\
$|\Delta\phi_{\mu\mu} - \pi| > 0.8$ & $3.9\cdot 10^{5}\cdot{{\rm{BR}}_{\rm inv}^{\rm BSM}}$ & $1.7\cdot 10^{5}$   & $2.2\cdot 10^{6}$ & $4.9\cdot 10^{5}$ & $8.5\cdot 10^{4}$ & $4.6\cdot 10^{2}$\\
$P_{\bot}^{\mu\mu} > 80$~GeV & $3.4\cdot 10^{5}\cdot{{\rm{BR}}_{\rm inv}^{\rm BSM}}$ & $1.1\cdot 10^{5}$ & $8.9\cdot 10^{5}$ & $2.5\cdot 10^{5}$ &  $5.9\cdot 10^{4}$  & $4.1\cdot 10^{2}$\\
$M_{\mu\mu} > 9.5$~TeV & $1.6\cdot 10^{5}\cdot{{\rm{BR}}_{\rm inv}^{\rm BSM}}$ & $1.4\cdot 10^{3}$ & $1.5\cdot 10^{3}$ & $3.1\cdot 10^{2}$ & $28$  &  $1.9\cdot 10^{2}$ \\
$E_{\mu}^{\rm min} > 4.7$~TeV & $1.1\cdot 10^{5}\cdot{{\rm{BR}}_{\rm inv}^{\rm BSM}}$ & $6.2\cdot 10^{2}$ & $(< 65)$ & $(< 31)$ &  $(< 7.0)$  & $1.3\cdot 10^{2}$
\\\hline
\end{tabular}\caption{Cut-flow for $2E_{\rm b} = 10$~TeV and a forward detector coverage $\eta_\mu^{\rm max} = 6$. An energy smearing of $1\%$ is applied to muons with $|\eta_\mu| > 2.44$. The baseline cuts are listed in Section~\ref{sec:truth}. 
}
\label{tab:cut_flow_1p}
\end{table*}

\renewcommand{\tabcolsep}{6pt}
\renewcommand{\arraystretch}{1.4}
\begin{table*}[t]
\centering
\begin{tabular}{ lcccccc }    
\hfil [number of events, $10$~ab$^{-1}$] & BSM signal &  $\mu^+\mu^-\bar{\nu}\nu$ &  $\mu^+\mu^- \gamma$ & $\mu^+\mu^- \bar{f} f$ & $\mu^+\mu^- W^+ W^-$  & $\mu^+\mu^- (h\rightarrow \text{inv})_{\rm SM}$\\\hline
baseline $\&\, P_{\bot}^{\mu\mu} > 50$~GeV & $6.2\cdot 10^{5}\cdot{{\rm{BR}}_{\rm inv}^{\rm BSM}}$ & $1.1\cdot 10^{6}$  & $1.5\cdot 10^{7}$ & $1.3\cdot 10^{6}$ &  $6.2\cdot 10^{5}$  & $7.4\cdot 10^{2}$\\
$|\Delta \eta_{\mu\mu}| > 6.5$ & $6.1\cdot 10^{5}\cdot{{\rm{BR}}_{\rm inv}^{\rm BSM}}$ & $7.6\cdot 10^{5}$   & $1.3\cdot 10^{7}$ & $1.1\cdot 10^{6}$ & $5.5\cdot 10^{5}$  & $7.3\cdot 10^{2}$\\
$|\Delta\phi_{\mu\mu} - \pi| > 1$ & $4.4\cdot 10^{5}\cdot{{\rm{BR}}_{\rm inv}^{\rm BSM}}$ & $3.9\cdot 10^{5}$   & $2.9\cdot 10^{6}$ & $6.4\cdot 10^{5}$ & $3.0\cdot 10^{5}$ & $5.3\cdot 10^{2}$\\
$P_{\bot}^{\mu\mu} > 180$~GeV & $1.9\cdot 10^{5}\cdot{{\rm{BR}}_{\rm inv}^{\rm BSM}}$ & $1.1\cdot 10^{5}$ & $2.7\cdot 10^{5}$ & $8.2\cdot 10^{4}$ &  $7.0\cdot 10^{4}$  & $2.2\cdot 10^{2}$\\
$M_{\mu\mu} > 8.75$~TeV & $1.2\cdot 10^{5}\cdot{{\rm{BR}}_{\rm inv}^{\rm BSM}}$ & $4.4\cdot 10^{3}$ & $7.6\cdot 10^{3}$ & $1.9\cdot 10^{3}$ & $1.6\cdot 10^{3}$  &  $1.4\cdot 10^{2}$ \\
$E_{\mu}^{\rm min} > 4.3$~TeV & $8.1\cdot10^{4}\cdot{{\rm{BR}}_{\rm inv}^{\rm BSM}}$ & $1.8\cdot 10^{3}$ & $2.6\cdot 10^{2}$ & $1.6\cdot 10^{2}$ &  $2.6\cdot 10^{2}$  & $97$
\\\hline
\end{tabular}\caption{Cut-flow for $2E_{\rm b} = 10$~TeV and a forward detector coverage $\eta_\mu^{\rm max} = 6$. An energy smearing of $10\%$ is applied to muons with $|\eta_\mu| > 2.44$. The baseline cuts are listed in Section~\ref{sec:truth}.}
\label{tab:cut_flow_10p}
\end{table*}

An energy resolution $\delta_{\rm res} = 1\%$ is most likely unrealistic, and larger uncertainties will deteriorate the reach significantly, limiting the muon collider sensitivity above the $10^{-4}$ level. This is illustrated by our results for $\delta_{\rm res} = 10\%$. With this uncertainty, the MIM is no longer a useful discriminant. The $E_\mu^{\rm min}$ distributions of both signal and backgrounds extend beyond the truth-level endpoint $E_{\rm b}$ (see Fig.~\ref{fig:smeared}), significantly reducing the rejection power of this variable. Similar considerations hold for $M_{\mu\mu}$. The optimal sensitivity is obtained through softer cuts than in the $\delta_{\rm res} = 1\%$ scenario, as displayed by the cut-flow in Table~\ref{tab:cut_flow_10p}. The SM branching ratio produces 100 events in the selected region, with more than 2000 background events. It should be possible to observe the SM Higgs to invisible decay, but only at the $95\%$ confidence level. A $5\sigma$ ``discovery'' of the SM invisible decay could be impossible. The sensitivity to new physics is $\mathrm{BR}_{\rm inv}^{\rm BSM} < 10^{-3}$ as in Fig.~\ref{fig:BR10TeV}. At the exclusion, $\mathrm{S}/\mathrm{B}\approx 3\%$.

It should be noted that our sensitivity projections based on cut-and-count could be improved by a more sophisticated statistical analysis. On the other hand, these possible improvements are not expected to modify the picture radically. 

\begin{figure}[t]
\includegraphics[width=0.48\textwidth]{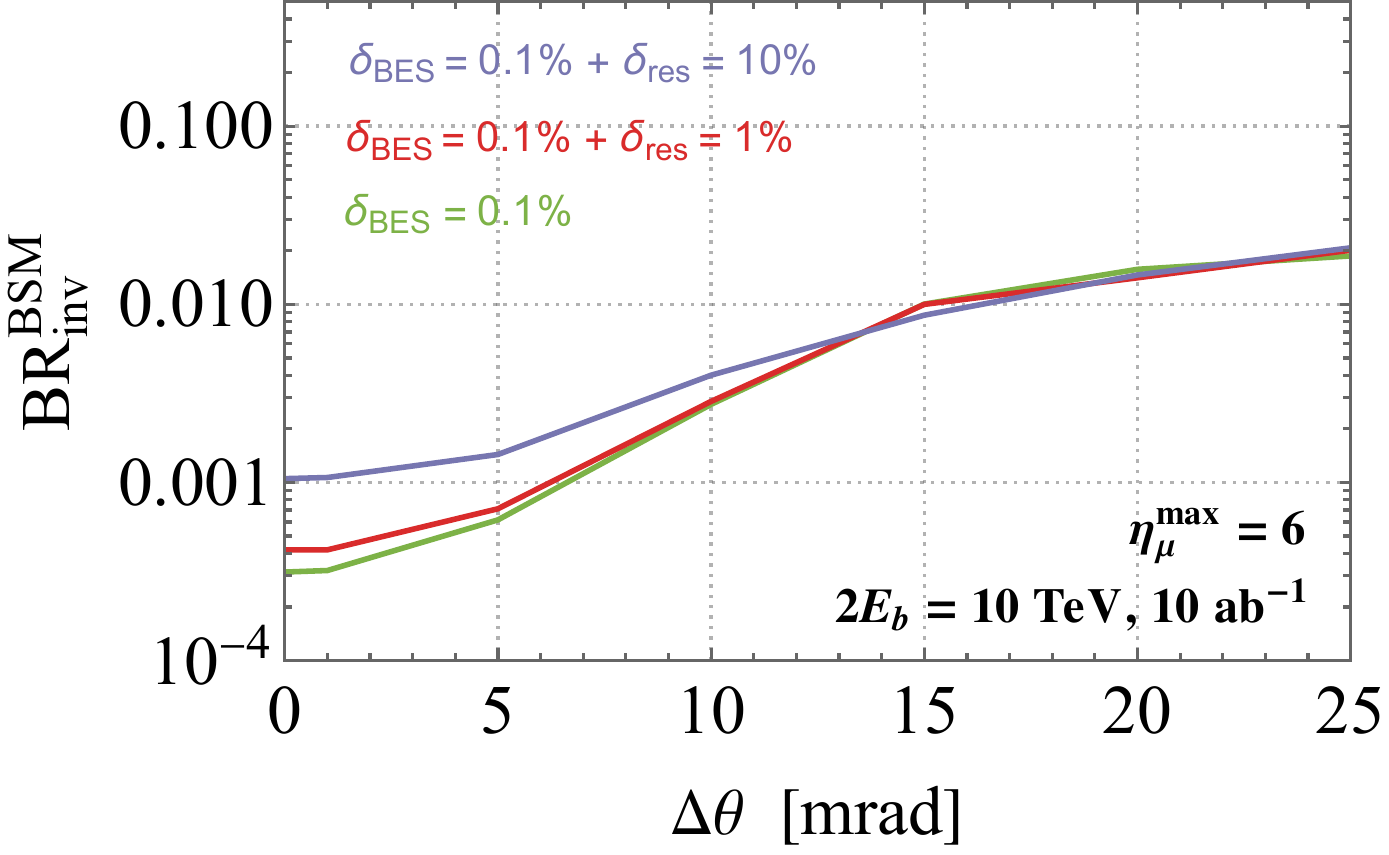}
\caption{\label{fig:BR10TeVAngSmear} 95\% CL limit on BR$_{\rm inv}^{\rm BSM}$ at a 10~TeV muon collider with $10$~ab$^{-1}$, as a function of the uncertainty on the muon direction measurement. An angular acceptance $\eta_\mu^{\rm{max}}=6$ of the forward muon detector is assumed.}
\end{figure}

We finally turn to the investigation of the effect of uncertainties in the muon direction measurement. The sensitivity is shown in Fig.~\ref{fig:BR10TeVAngSmear} as a function of the angular uncertainty $\Delta\theta$ assuming a forward detector coverage up to $\eta_\mu^{\rm max} = 6$. For $1\%$ energy resolution, an angular resolution $\Delta\theta<2$~mrad would be needed not to affect the sensitivity. If instead $\delta_{\rm res} = 10\%$, the sensitivity is inferior and it does not get degraded significantly up to $\Delta\theta=5$~mrad. Larger $\Delta\theta$, above around 10~mrad, would most likely prevent the observation of the SM decay regardless of the energy measurement accuracy. 

\section{Conclusions}\label{sec:conc}

The installation of a forward muon detector would further improve the perspectives to detect and to investigate vector boson fusion or scattering processes at a muon collider. An extensive survey of specific opportunities offered by the forward muon detector will be presented elsewhere~\cite{toappear}. In this paper we studied the Higgs boson decay to invisible particles.

Our main results are displayed in Figs.~\ref{fig:BR10TeV} and~\ref{fig:BR10TeVAngSmear}. A sensitivity to $\mathrm{BR}_{\rm inv}^{\rm BSM}$ as small as $10^{-4}$, competitive with the FCC-hh projection, is ideally possible  assuming perfect discrimination of the $\mu^+ \mu^- (h\to{\rm{inv}})$ process from the backgrounds. We have shown that nearly perfect discrimination is in principle possible, and the ideal sensitivity is attainable, by fully realistic simulations of the underlying physical processes that also account for the energy and angular spread of the colliding muon beams. 

However, a strong sensitivity degradation is expected due to the finite resolution in the measurement of the energy and of the angle of the muons in the forward detector. An energy resolution as small as $1\%$ and an angular uncertainty below 2~mrad would be needed to maintain a sensitivity at the $10^{-4}$ level. With more realistic resolutions, such as for instance $10\%$ on the energy, the sensitivity drops to the $10^{-3}$ level and even the possibility of observing the SM Higgs to invisible decay cannot be taken for granted. Our results motivate design studies of the forward muon detector targeting the best possible accuracy on the muon momentum measurement. There could be margins to improve the sensitivity by multivariate shape analyses that should be also investigated.

Our findings rely on assumptions on the performances of the main detector. In particular, we assumed coverage up to $\theta=10^{\rm o}$ from the beam axis, and 20~GeV threshold for the veto of any object within the angular acceptance. The validity of these assumptions cannot be verified at the current stage, because the design of the 10~TeV muon collider detector has not been completed yet. We show in Appendix~\ref{app:main} (see Fig.~\ref{fig:BR10TeVMainDet}) that our findings depend weakly on them.

A first stage of the muon collider project could foresee a 3~TeV centre of mass energy collider with 2~ab$^{-1}$ integrated luminosity~\cite{Accettura:2023ked}. The projected sensitivity of a 3~TeV muon collider is studied in Appendix~\ref{app:3}. We find, in Fig.~\ref{fig:BR3TeV}, that observing the SM Higgs to invisible decay will most likely be impossible for $\delta_{\rm res}=10\%$. The reduced prospects are due to the smaller number of produced Higgs bosons. On the other hand, the  3~TeV muon collider will still improve over the HL-LHC sensitivity. An angular coverage of the forward muon detector up to 4 or 5 pseudo-rapidity would be sufficient, while coverage up to 5 or 6 is needed for the 10~TeV collider.

\begin{acknowledgments}
We thank Daniele Calzolari, Riccardo Masarotti, Federico Meloni, Simone Pagan Griso, J\"urgen Reuter and Daniel Schulte for useful discussions. MR and AW thank the KITP, which is supported by the National Science Foundation under Grant No.~NSF PHY-1748958, where part of this work was completed. ES thanks the IFAE at the UAB for hospitality in the final stages of the project. MR is supported by the NSF grant PHY-2014071 and by a Feodor--Lynen Research Fellowship awarded by the Humboldt Foundation. ES acknowledges partial support from the EU’s Horizon 2020 programme under the MSCA grant agreement 860881-HIDDeN. AW is supported by the grant PID2020-115845GB-I00/AEI/10.13039/501100011033. 
\end{acknowledgments}

\appendix

\section{Impact of angular coverage and veto threshold of the main detector}\label{app:main}

Here we study the sensitivity of our results to variations of the angular coverage and of the veto $p_\bot$ threshold of the main detector. Specifically, we consider the possibility that the size of the absorbers is reduced with respect to current design, thus freeing space to extend the angular coverage up to $\theta = 5^{\rm o}$ ($|\eta| \approx 3.1$) instead of the $10^{\rm o}$ assumed in the main text. We also investigate the effect of increasing the veto $p_\bot$ threshold to 50~GeV from the 20~GeV considered so far.

\begin{figure}[t]
\includegraphics[width=0.48\textwidth]{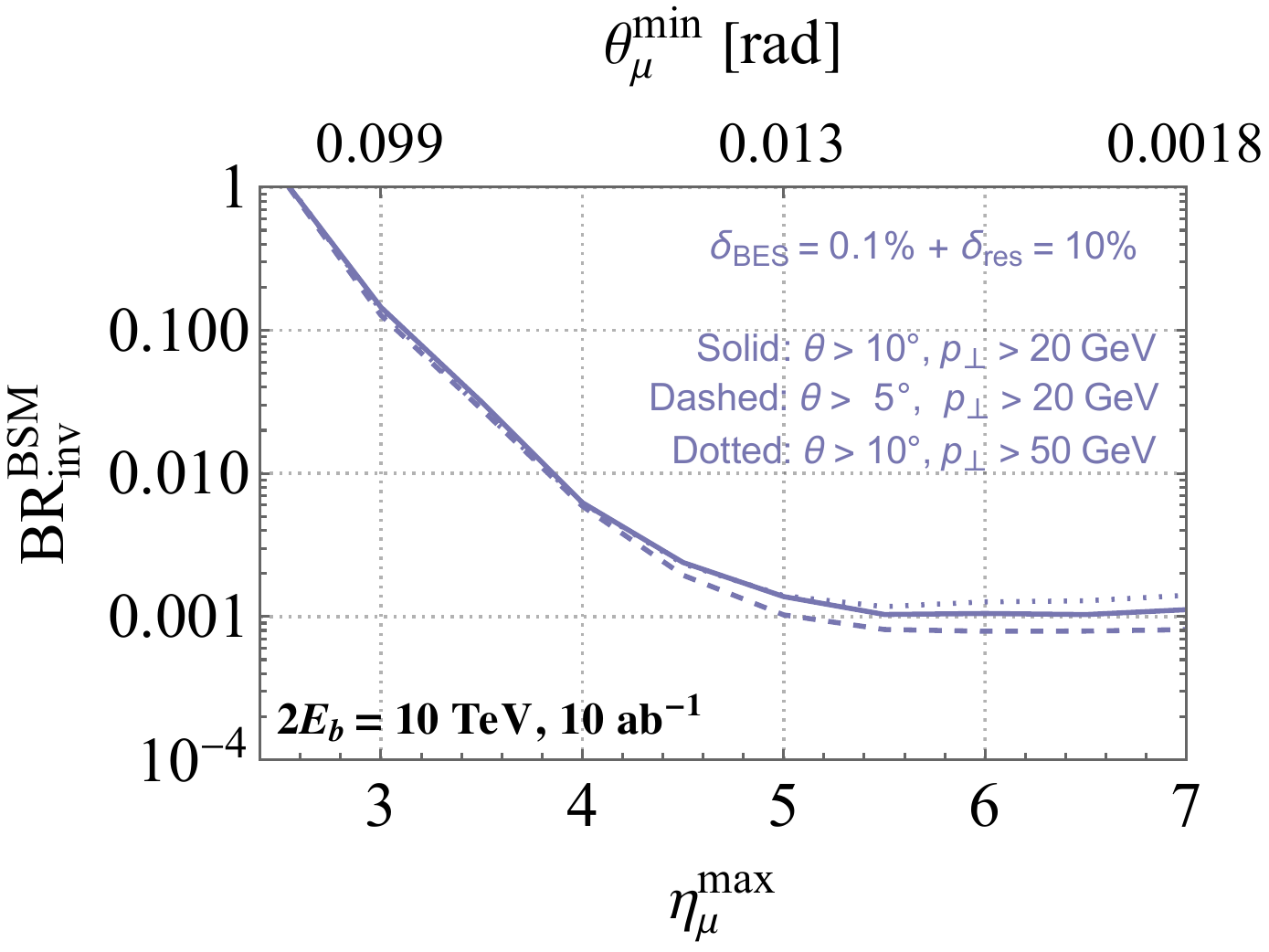}
\caption{\label{fig:BR10TeVMainDet} 95\% CL limit on BR$_{\rm inv}^{\rm BSM}$ at a 10~TeV muon collider with an extended angular coverage up to $\theta = 5^\circ$ (dashed) and a larger energy threshold (dotted) for particle detection of the main detector. A 10\% energy measurement uncertainty for forward muons, and no angular uncertainty, is included.}
\end{figure}

The major effect of extending the angular coverage of the main detector is a better rejection of background processes where the $\mu^+\mu^-$ pair is produced together with an additional object at moderate pseudo-rapidity. These include the $\mu^+\mu^- W^+ W^-$ and $\mu^+\mu^- f\bar{f}$ processes, for which we find significant suppression. The $\mu^+\mu^- \gamma$ background is mildly reduced as well. Even though the emitted photons are preferentially collinear with the initial or final state muons and thus very forward, the $P_{\bot}^{\mu\mu}$ cut applied in our analysis eventually selects events where the photon has larger $p_\bot$ and is therefore more central.
Quantitatively, assuming a forward detector coverage of $\eta_\mu^{\rm max} = 6$ and after the combination of baseline cuts and $P_{\bot}^{\mu\mu} > 50$~GeV, the reduction amounts to 65\%, 40\% and 13\% for the $\mu^+\mu^- W^+ W^-$, $\mu^+\mu^-  f\bar{f}$, and $\mu^+\mu^- \gamma$ backgrounds, respectively. The signal is not affected. This results in a mild improvement of the sensitivity to BR$_{\rm inv}^{\rm BSM} < 8\cdot 10^{-4}$ for $\delta_{\rm res} = 10\%$ and $\eta_\mu^{\rm max} = 6$, as shown by the dashed line in Fig.~\ref{fig:BR10TeVMainDet}. This weak dependence on the angular acceptance of the main detector can be understood by noticing that the dominant background after all cuts is $\mu^+\mu^-\bar{\nu}\nu$, which is not affected by the veto. 


If the veto $p_\bot$ threshold is larger than the $20$~GeV assumed in the main text, the rejection of background events with soft particles in the central region becomes less effective. We find that increasing the threshold to $p_\bot > 50$~GeV (while keeping the angular coverage of the main detector fixed to $\theta > 10^\circ$) has a very mild effect. The $\mu^+\mu^- W^+ W^-$ and $\mu^+\mu^-  f\bar{f}$ backgrounds are larger by 25\% after baseline cuts and $P_{\bot}^{\mu\mu} > 50$~GeV, whereas all remaining backgrounds, including the dominant $\mu^+\mu^-\bar{\nu}\nu$ process, and the signal are approximately unchanged. This small increase in the background rate results in a modest 10\% degradation of the sensitivity to BR$_{\rm inv}^{\rm BSM}$, displayed by the dotted line in Fig.~\ref{fig:BR10TeVMainDet}.

We conclude that the design of the main detector has far milder impact on the sensitivity to BR$_{\rm inv}^{\rm BSM}$ compared to the energy and angular resolution of the forward muon detector.

\section{The 3~TeV muon collider}\label{app:3}

In this Appendix we study the sensitivity of a muon collider at a centre of mass energy of 3~TeV and integrated luminosity of 2~ab$^{-1}$ to invisible Higgs decays using forward muons. The generation of MC data samples and the analysis are analogous to the 10~TeV study discussed in the main text, including the simulation of the finite energy resolution of the forward detector. However, we do not consider the uncertainties in the measurement of the muon direction. At 3~TeV the muons produced by the signal process have smaller pseudo-rapidity, with a distribution peaking at $\theta \approx 61$~mrad ($|\eta| \approx 3.5$). This implies that muons inside the acceptance of the main detector, taken to be $\theta > 10^{\rm o}$ here, play a non-negligible role. Therefore, when a finite energy resolution $\delta_{\rm res}$ is assumed for the forward detector we also apply a $1\%$ energy smearing to muons with $|\eta| < 2.44$. 

\begin{figure}[b]
\includegraphics[width=0.48\textwidth]{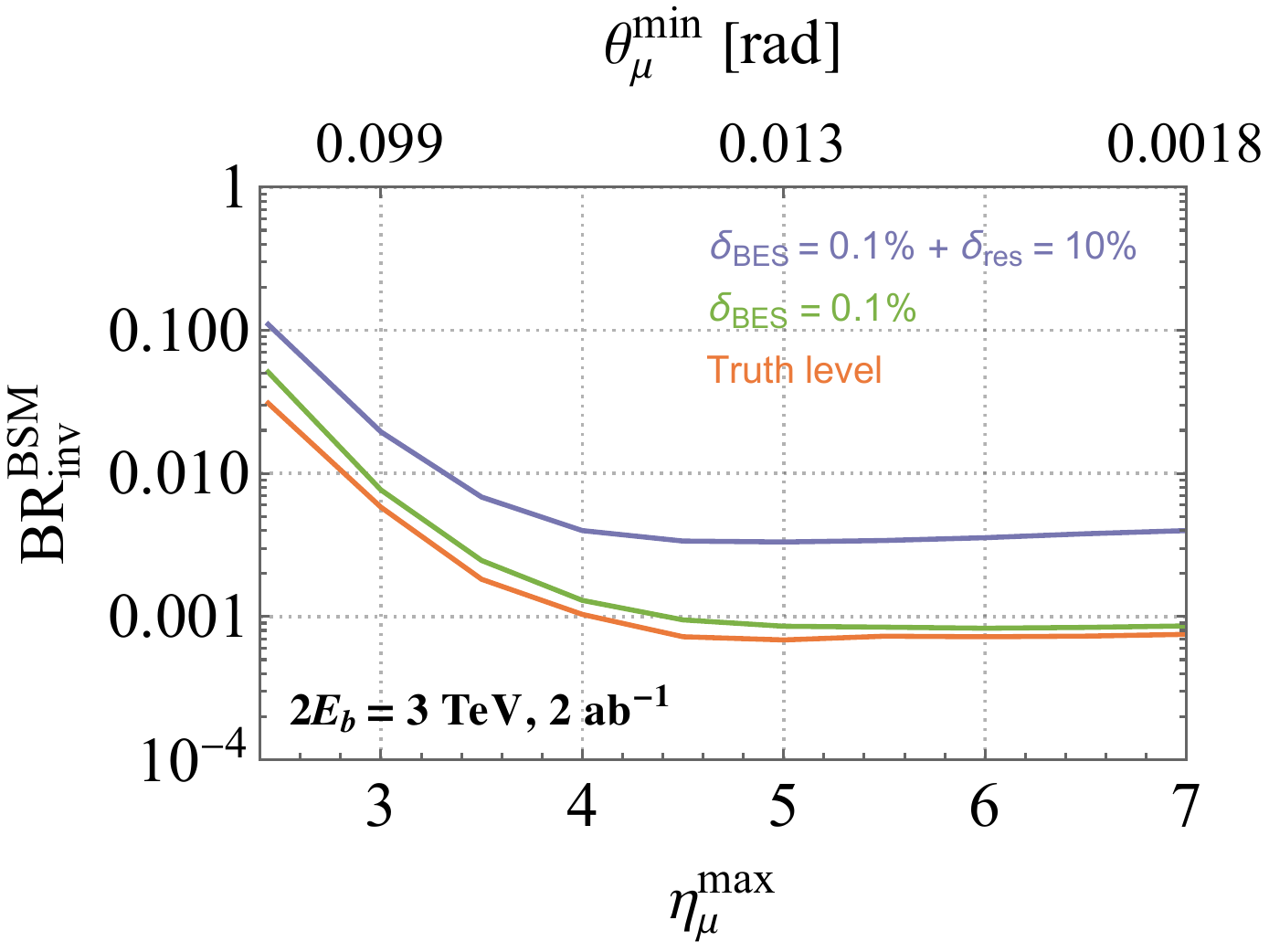}
\caption{\label{fig:BR3TeV} 95\% CL limit on BR$_{\rm inv}^{\rm BSM}$ at a 3~TeV muon collider with $2$~ab$^{-1}$, as a function of the angular acceptance $2.44 < |\eta_\mu| < \eta_\mu^{\rm max}$ of a forward muon detector. The main detector is assumed to cover $\theta > 10^{\rm o}$.}
\end{figure}

The expected reach is shown in Fig.~\ref{fig:BR3TeV}, demonstrating that a smaller angular coverage compared to the 10~TeV case, up to $\eta_\mu^{\rm max} \approx 4\,$-$\,5$, would be sufficient to obtain optimal sensitivity. The limits on $\mathrm{BR}_{\rm inv}^{\rm SM}$ are weaker by a factor of a few compared to a 10~TeV collider owing to the much lower number of produced Higgs bosons. For instance, assuming angular coverage $\eta_\mu^{\rm max} = 5$ and energy resolution $\delta_{\rm res} = 10\%$ we find BR$_{\rm inv}^{\rm BSM} < 3\cdot 10^{-3}$. These results suggest it is unlikely that the SM Higgs invisible decay can be observed at a 3~TeV muon collider, but a decisive sensitivity improvement relative to HL-LHC is expected.

\bibliography{bibliography}

\end{document}